\documentclass[a4paper,fleqn,usenatbib]{mnras}

\usepackage[T1]{fontenc}
\usepackage{ae,aecompl}

\usepackage{graphicx}   
\usepackage{amsmath}    
\usepackage{bm}       
\usepackage{amssymb}    
\usepackage{wasysym}  
\usepackage{txfonts}
\usepackage{etoolbox}
\makeatletter 
  \patchcmd{\NAT@citex}
    {\@citea\NAT@hyper@{%
      \NAT@nmfmt{\NAT@nm}%
      \hyper@natlinkbreak{\NAT@aysep\NAT@spacechar}{\@citeb\@extra@b@citeb}%
      \NAT@date}}
    {\@citea\NAT@nmfmt{\NAT@nm}%
    \NAT@aysep\NAT@spacechar\NAT@hyper@{\NAT@date}}{}{}

  \patchcmd{\NAT@citex}
    {\@citea\NAT@hyper@{%
      \NAT@nmfmt{\NAT@nm}%
      \hyper@natlinkbreak{\NAT@spacechar\NAT@@open\if*#1*\else#1\NAT@spacechar\fi}%
        {\@citeb\@extra@b@citeb}%
      \NAT@date}}
    {\@citea\NAT@nmfmt{\NAT@nm}%
    \NAT@spacechar\NAT@@open\if*#1*\else#1\NAT@spacechar\fi\NAT@hyper@{\NAT@date}}
    {}{}
\makeatother

\newcommand\Msun{\text{M}_{\astrosun}} 
\newcommand\colt{\textsc{colt}} 
\newcommand\HI{{H\,\textsc{i}}} 
\newcommand\HeII{{He\,\textsc{ii}}} 
\newcommand\OIII{{O\,\textsc{iii}}} 

\usepackage[multiple]{footmisc}
\makeatletter
\renewcommand\@makefntext[1]%
     {\textsuperscript{\@thefnmark}\,#1}
\makeatother
\newcommand\ASfootnotemark[1][1]{\textsuperscript{\footnotemark[#1]}}

\title[The first supermassive black holes]{%
  \vspace{-1em}
  \includegraphics[width=\textwidth]{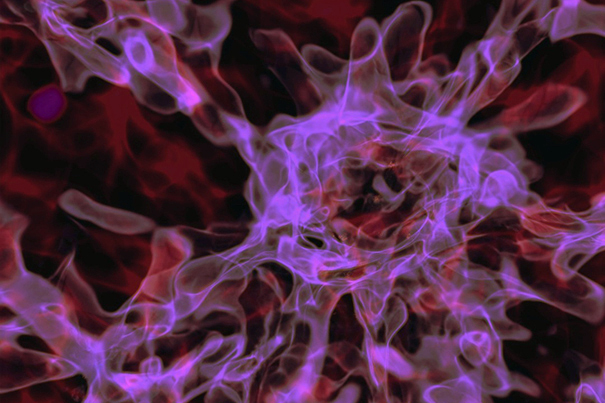} \\
  \vspace{.75em}
  The first supermassive black holes
}

\author[A.\ Smith et al.]{
  Aaron~Smith,$^1$ 
  Volker~Bromm$^1$ and
  Abraham~Loeb$^2$
  \\
  $^1$Department of Astronomy, The University of Texas at Austin, Austin, TX 78712, USA \\
  $^2$Department of Astronomy, Harvard University, 60 Garden Street, Cambridge, MA 02138, USA
}

\date{\vspace{-3em}}

\pubyear{2017}

\begin{document}
\label{firstpage}
\pagerange{\pageref{firstpage}--\pageref{lastpage}}
\maketitle


The existence of a supermassive black hole (SMBH) at the centre of almost every galaxy provides strong evidence for a generic growth mechanism alongside that of the host galaxy \citep{Kormendy_2013}. Nuclear black holes typically comprise very much less than 1\% of the available baryonic mass in the galaxy and are therefore gravitationally dominant only locally. But their efficient conversion of rest mass to radiation plays an important role in the evolution of galaxies as a powerful source of feedback that drives winds, suppresses star formation and contributes to the reionization of the universe. Understanding the physical processes leading to the formation and growth of SMBHs is crucial in unraveling fundamental questions about the galactic building blocks of the universe. While the largest-scale structures took several billion years to form, supermassive black holes up to $\sim 10^{10}\,\Msun$ (quasars) are observed very early in the history of the universe, around a cosmological redshift of $z \approx 7$, at less than $5$\% of the current age of the universe \citep{Fan_2006,Mortlock_2011,Wu_2015}. This presents a timing problem under conventional scenarios for the growth of these objects by accretion.

The rate of black hole growth via gas consumption is regulated by balancing the attractive gravitational force against the build-up of pressure from the radiation emitted by the infalling gas as it encounters viscous heating. For typical radiative efficiencies, the maximal growth rate -- known as the Eddington limit -- 
barely allows growth from stellar-mass black hole progenitors in the limited time available since the Big Bang. As an analogy, we consider the exponential growth of a bank account accruing compound interest. A lifetime of steady investments makes the burden of saving for retirement relatively straightforward and secure. However, if the growth has to occur in less than a decade then there is a timing crisis: the unfortunate individual must either save money at exorbitant rates, experience a miracle akin to winning the lottery -- or delay their retirement.

\subsection*{Timing Matters}
\begin{figure}
    \centering
    \includegraphics[width=\columnwidth]{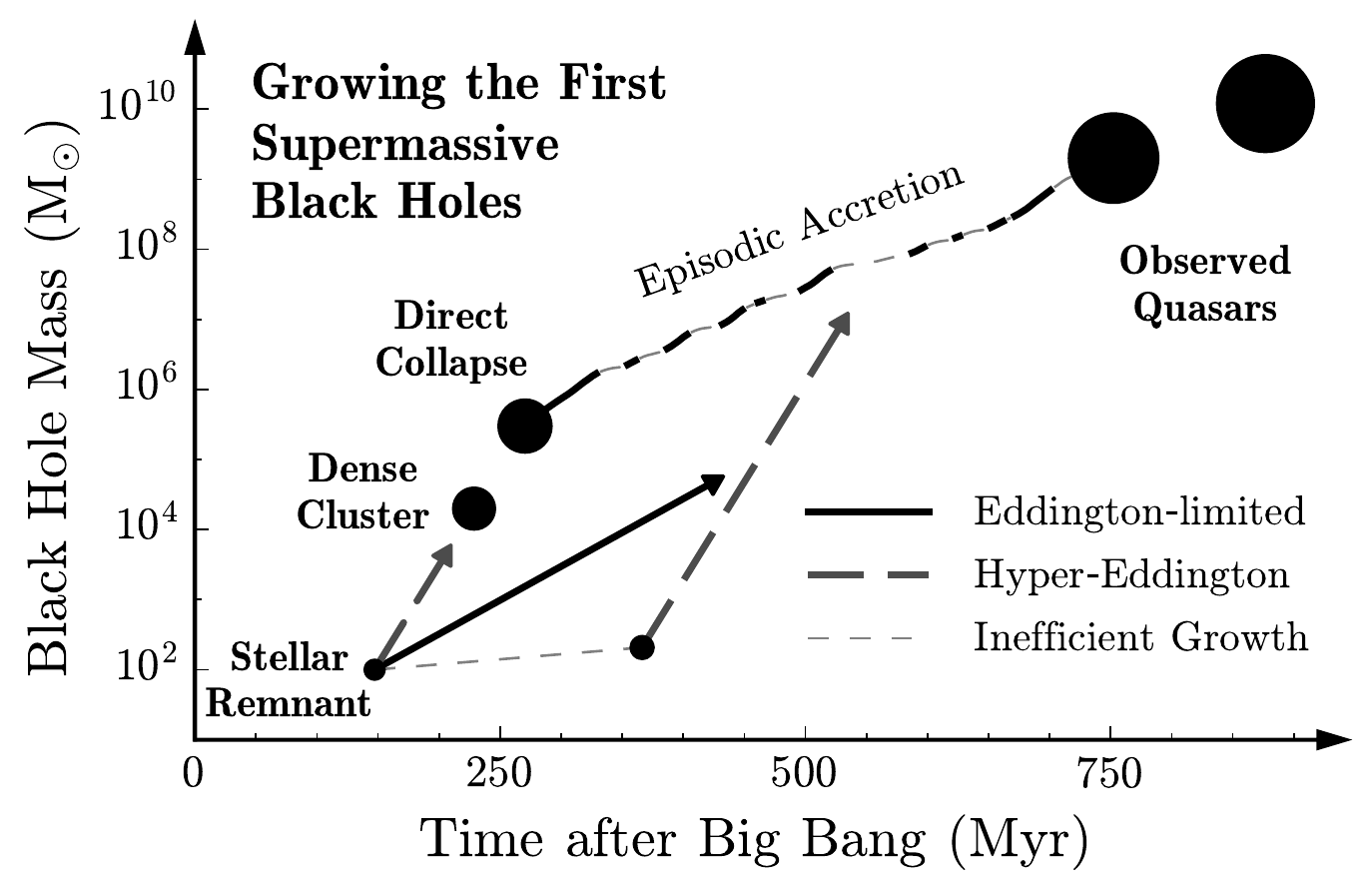}
    \caption{\protect\input{figures/M_BH/caption}}
    \label{fig:M_BH}
  \end{figure}

How and when did the first supermassive black holes form? Although there may not be a universal pathway, rapid seeding or growth is unavoidable within our current understanding of the most distant quasars. The first SMBHs were likely to have co-assembled within the first galaxies a few hundred million years after the Big Bang. Remarkably, primordial galaxies that could delay the onset of star formation seem to facilitate a mechanism to produce massive ($10^4$--$10^6\,\Msun$) black hole seeds, which form ``in one go'' from gas clouds with inefficient cooling. Such clouds thereby maintain high thermal pressure support, which in turn suppresses fragmentation and thus star formation. These direct collapse black holes (DCBHs) circumvent the timing crisis. DCBHs can form only early in cosmic history when a specific set of rare conditions is satisfied. At redshifts $z \gtrsim 10$, they are still beyond the reach of current telescopes. Progress will be made with next-generation observatories using wide-field surveys and follow-up deep observations of select objects. Ultimately, this will provide compelling evidence for or against formation scenarios involving direct-collapse, persistent hyper-Eddington accretion, or other more exotic possibilities \citep[e.g.][]{Dolgov_Silk_1993}. In the sections below, we briefly review the historical development of ideas about the first SMBH seeds, the physics of their formation and radiative feedback, recent progress, and our outlook for the future. Figure~\ref{fig:M_BH} provides a schematic overview of the quasar-seed timing problem.

\subsection*{Cosmological context}
The standard cosmological model involves dark energy and cold dark matter ($\Lambda$CDM). Both ingredients remain mysterious in terms of their underlying physics. However, the $\Lambda$CDM model matches data across scales extending from large-scale structures to galaxy formation and evolution. Historically, supermassive black hole research predates the recent cosmological perspective. Summarizing these earlier insights, \citet{Rees_1984} discussed possible routes for runaway growth in active galactic nuclei. In isolated environments, possible formation pathways include: stellar-remnant black holes after vigorous gas accretion; dense star clusters in which runaway collisions trigger the formation of a SMBH; or a cluster of post-supernovae neutron stars or black holes coalescing in the dynamically unstable central core of the galaxy. In cosmology, SMBHs can grow over longer timescales via episodic galaxy mergers and accretion from streams of cold gas along filaments of the cosmic web \citep{Mayer_2010,Mayer_2015}. Still, it was recognized that a gas cloud may conceivably bypass conventional star formation and yield a near-extremal Kerr black hole if cooling does not initiate fragmentation \citep{Rees_1984}.

Early simulations of collapsing primordial gas clouds showed that most of the gas does indeed fragment into dense stellar clumps which eventually virialize into a spheroidal galactic bulge \citep{Loeb_Rasio_1994}. The emergence of a SMBH would thus be forestalled. On the other hand, if a central seed black hole of mass $\gtrsim 10^6~\Msun$ were to form during the collapse then it could quickly grow by steady accretion to a quasar-size black hole \citep{Li_2007}. Otherwise dynamical instabilities in the inner region of the disc would inhibit the accretion required for such growth. The crucial bottleneck is the formation of the initial, million solar-mass, seed. Either way, realistic modeling of massive black hole inception is highly complex and ultimately relies on understanding both small- and large-scale phenomena. For example, low angular momentum configurations would help facilitate runaway collapse because the centrifugal barrier is significantly lower; in this context, quasar seeds could be a natural consequence of the initial collapse of regions with unusually small rotation \citep{Eisenstein_Loeb_1995}. Such low-spin cosmological perturbations might provide environments in which SMBH formation is an extreme manifestation of the $\Lambda$CDM model.

\subsection*{Forming the first massive black holes}
The main contenders for the earliest quasar seeds are DCBHs, super-Eddington accretion onto stellar remnant black holes, and runaway collisions in dense star clusters. Here we focus on DCBHs and highlight recent theoretical and observational evidence for this new class of black hole seeds; we discuss the alternative scenarios briefly.

In typical galactic environments black hole accretion is episodic because of self-regulating radiative feedback which yields accretion rates that are sub-Eddington when averaged over multiple duty cycles \citep{Johnson_2007,Milosavljevic_2009}. However, maintaining super-Eddington accretion is possible when the black hole is embedded within sufficiently dense gas; this renders the radiation pressure less effective \citep[e.g.][]{Wyithe_Loeb_2012,Pacucci_2015}. Based on one-dimensional radiation-hydrodynamics simulations, \citet{Inayoshi_2016} find accretion rates exceeding $\dot{M}_\bullet \gtrsim 10^3~L_\text{Edd}/c^2$ when the following condition is satisfied: $(n_\infty/10^5\,\text{cm}^{-3}) > (M_\bullet/10^4\,\Msun)^{-1}\;(T_\infty/10^4\,\text{K})^{3/2}$, where $n_\infty$ and $T_\infty$ are the density and temperature of the ambient gas. It remains an open question whether such growth rates are sustainable considering the violent assembly environments of the first galaxies where newly formed stars and supernovae blow away the surrounding gas. Other scenarios may also work, such as dense star clusters that undergo runaway collapse. With a ubiquitous supply of cold gas effectively trapping accretion radiation, a $\sim 10\,\Msun$ black hole seed undergoing random motions through the cluster may initiate supra-exponential growth over a dynamical timescale \citep{Alexander_2014}.

  \begin{figure}
    \centering
    \includegraphics[width=\columnwidth]{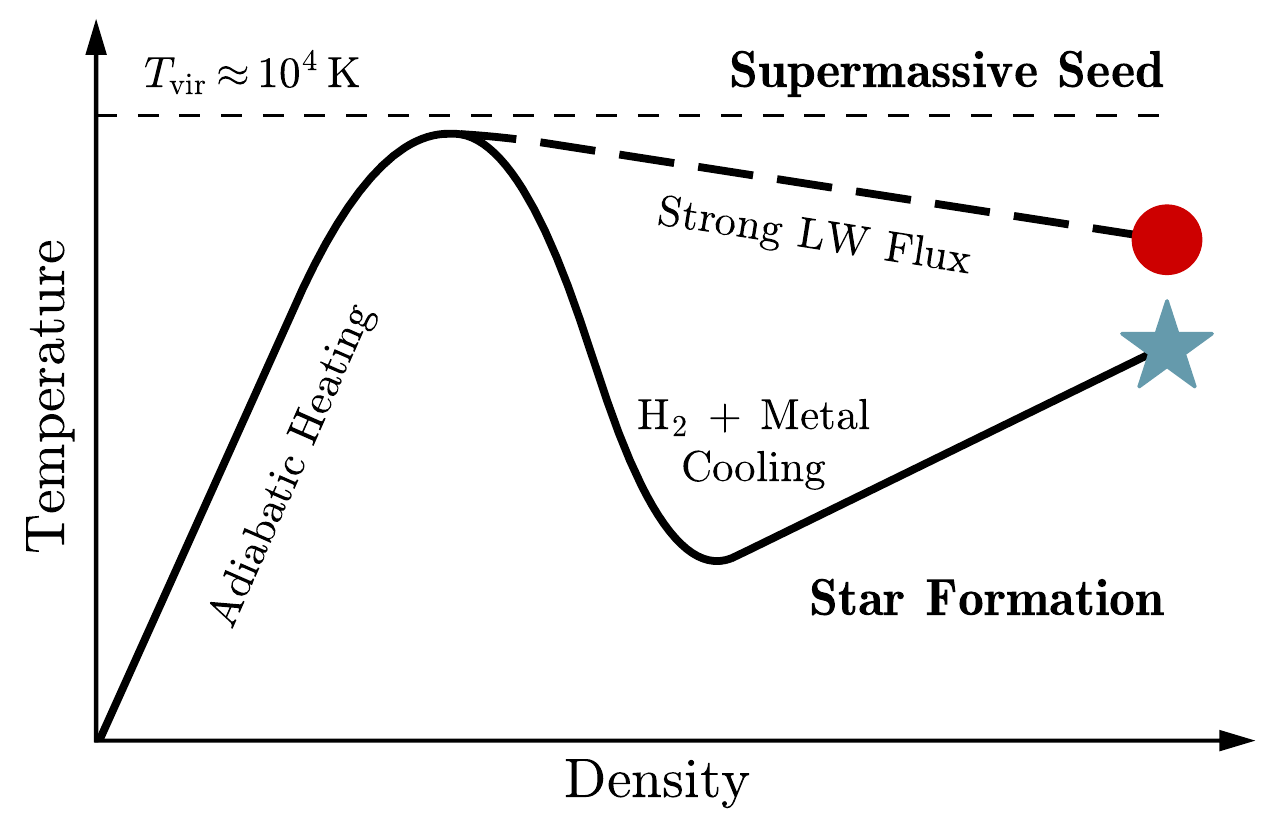}
    \caption{\protect\input{figures/cooling/caption}}
    \label{fig:cooling}
  \end{figure}

On the other hand, forming a DCBH requires collapse without fragmentation, a cosmic ``miracle'' of sorts, that is naturally explained within the context of galaxy formation theory. The idea is that if a primordial gas cloud, devoid of any heavy chemical elements (``metals'' in astronomical terminology), is bathed in a sea of ultraviolet radiation then it will be unable to cool and form stars (figure~\ref{fig:cooling}). Specifically, according to the Jeans criterion for triggering gravitational instability, high thermal pressure is required to prevent the gas cloud from fragmenting. In present-day star-forming clouds, line cooling by heavy elements and dust radiates away thermal energy that would otherwise provide stability. But in the early universe only hydrogen and helium were available to cool the gas. Thus, a small abundance of molecular hydrogen (H$_2$) played a key role as the primary cooling agent below the $\sim 8000$\,K accessible via atomic hydrogen line cooling. If strong non-ionizing Lyman-Werner radiation (LW; with photon energies below $13.6$\;eV) from neighbouring galaxies photodissociates H$_2$ then the evolutionary track through density-temperature phase space is significantly altered (see Fig.~\ref{fig:cooling}). Only atomic cooling primarily through the Lyman-$\alpha$ (Ly-$\alpha$) line of hydrogen efficiently keeps the collapsing core below $\sim 10^4$\,K. The first idealized models of isothermal collapse including H$_2$ photodissociation still predicted that the cloud breaks up at late stages despite the higher temperature track \citep{Omakai_2001}. Gas collapse may be entirely suppressed in the minihaloes, with virial masses of $M_\text{vir} \lesssim 10^6\,\Msun$, that are predicted to host the first stars, when the LW background flux is above a critical value. More massive host haloes, on the other hand, in general become self-shielding so that molecules can form again, eventually leading to star formation \citep{Oh_Haiman_2002}. However, \citet{Bromm_Loeb_2003}, carrying out the first simulations of this collapse with cosmological initial conditions, recognized that under conditions of unusually strong LW irradiation, the inflow would continue on its near-isothermal track. The resulting free-fall collapse of the atomically-cooling gas could then produce massive black holes directly. Subsequently, the DCBH model has received increased attention \citep[e.g.][]{Begelman_2006,Regan_2009,Choi_2013,Choi_2015}. This body of work is broadly reviewed in \citet{Volonteri_2012}, \citet{Haiman_2013}, \citet{Loeb_Furlanetto_2013}, \citet{Johnson_Haardt_2016}, and \citet{Latif_Ferrara_2016}.

\subsection*{Radiation from the first black holes}
Active galactic nuclei are conspicuous manifestations of gas accretion onto supermassive black holes, with luminosities exceeding the total starlight of their host galaxies. The accretion disc radiates broadband emission from optical to X-ray wavelengths with a peak in the UV. The characteristic blackbody temperature for the Eddington luminosity near the event horizon is $T_\text{Edd} \approx 5 \times 10^5\,\text{K}~(M_\bullet/10^8\,\Msun)^{-1/4}$ \citep{Rees_1984}. DCBHs are born in gas-rich environments and may be self-shielding to ionizing photons. Still, the gas remains transparent to X-rays and the ``Compton-thick'' spectrum retains the non-thermal tail contributed by Bremsstrahlung radiation and magneto-hydrodynamical (MHD) processes \citep{Pacucci_MBH_Spectra_2015}. In these environments, radiation pressure from the black hole along with concurrent star formation and nearby supernovae are likely to have had a substantial impact on the host galaxy \citep{Jeon_2012}. Lower mass minihaloes with shallower gravitational potential wells ($M_\text{vir} \lesssim 10^5$--$10^7\,\Msun$) would have been especially susceptible to radiative feedback, which potentially depleted the reservoir of gas needed to fuel black hole growth \citep{Whalen_2004,Wise_2012}. The viability of the DCBH mechanism also relies on the regulation of chemical feedback because molecular hydrogen and metal cooling induces fragmentation \citep[][and recall figure~\ref{fig:cooling}]{Hartwig_2016}. Therefore, the eventual DCBH formation sites must remain free from star formation during collapse. Furthermore, the emission of UV and X-ray photons promotes H$_2$ formation and increases the critical LW flux needed to form DCBHs \citep{Inayoshi_Tanaka_2015,Latif_Bovino_2015}.

  \begin{figure}
    \centering
    \includegraphics[width=\columnwidth]{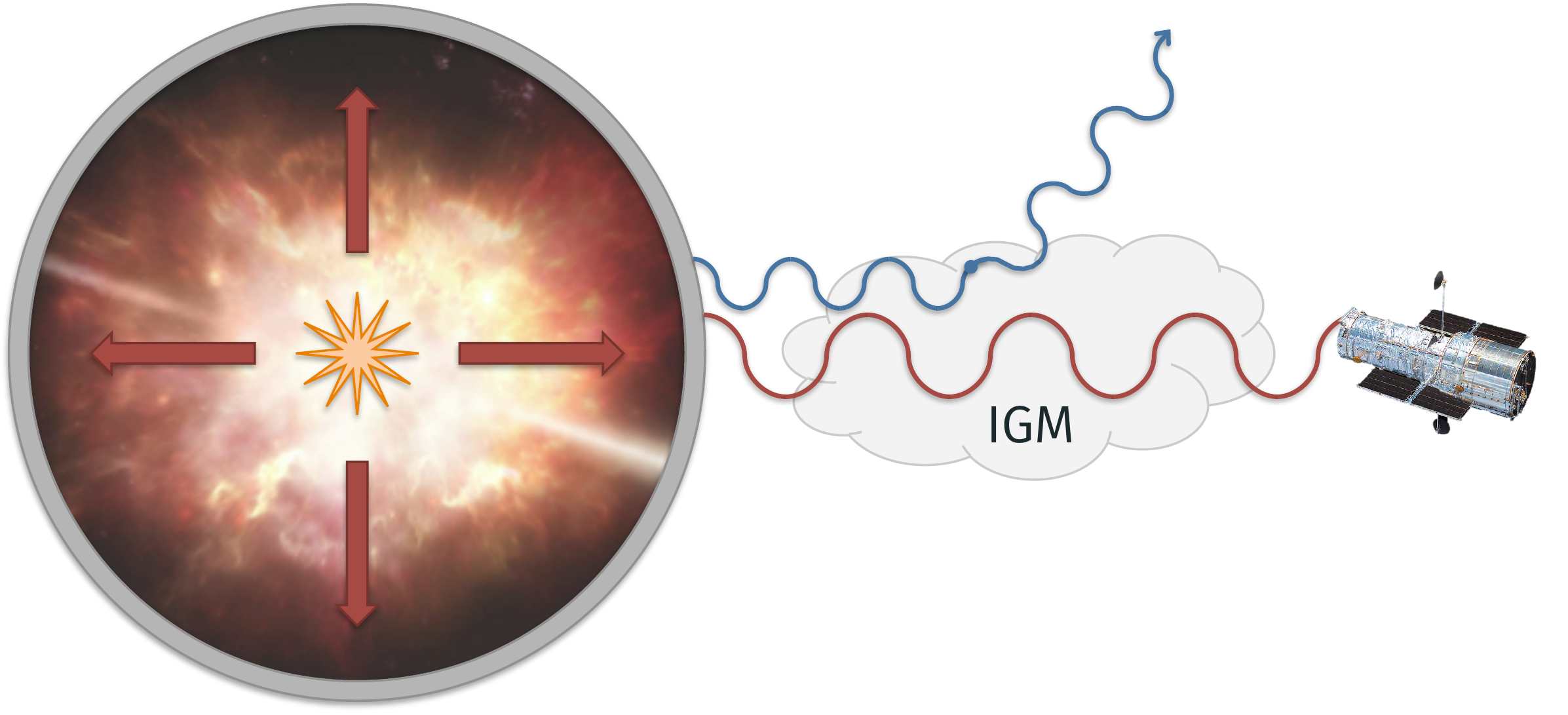}
    \caption{\protect\input{figures/CR7-cartoon/caption}}
    \label{fig:CR7-cartoon}
  \end{figure}

A particularly interesting source of feedback in DCBH environments is Ly-$\alpha$ photon trapping. In the vicinity of a newly collapsing primordial gas cloud, the surrounding neutral hydrogen is extremely optically thick to photons near the Ly-$\alpha$ resonant line. In one-dimensional simulations, a dense shell-like outflow structure forms in hydrodynamical response to the central source ionizing and heating of the gas. As illustrated in figure~\ref{fig:CR7-cartoon}, the Ly-$\alpha$ photons are redshifted by the galactic outflow, producing a velocity offset that may result in considerably less scattering out of the line of sight by the intervening intergalactic medium (IGM). The complex nature of the Ly-$\alpha$ radiative transfer means that explorations of Ly-$\alpha$ feedback tend to focus on order-of-magnitude estimates based on idealized calculations \citep{Oh_Haiman_2002,McKee_Tan_2008,Milosavljevic_2009}. With the aid of post-processing Monte-Carlo radiative transfer (MCRT) methods, \citet{Dijkstra_Loeb_2008} found that multiple scattering within high \HI\ column density shells is capable of enhancing the effective Ly-$\alpha$ force by one or two orders of magnitude. The first self-consistent Ly-$\alpha$ radiation hydrodynamics (RHD) simulations were performed by \citet{Smith_RHD_2017}, who coupled a MCRT code \citep[\colt;][]{Smith_2015} with spherically symmetric Lagrangian frame hydrodynamics including ionizing radiation, non-equilibrium chemistry and cooling, and self-gravity. They found that Ly-$\alpha$ radiation pressure may have a significant dynamical impact on gas surrounding DCBHs, with Ly-$\alpha$ signatures characterized by larger velocity offsets than stellar counterparts if, in both cases, the Ly-$\alpha$ spectra are shaped by radiation-driven winds. Finally, it has recently been suggested that trapped Ly-$\alpha$ cooling radiation may enhance the formation of DCBHs by accounting for the photodetachment of H$^{-}$ ions, precursors to H$_2$, by Ly-$\alpha$ photons during collapse \citep{Johnson_Dijkstra_2017}.

\subsection*{Observational evidence for DCBHs}
\begin{figure}
    \centering
    \includegraphics[width=\columnwidth]{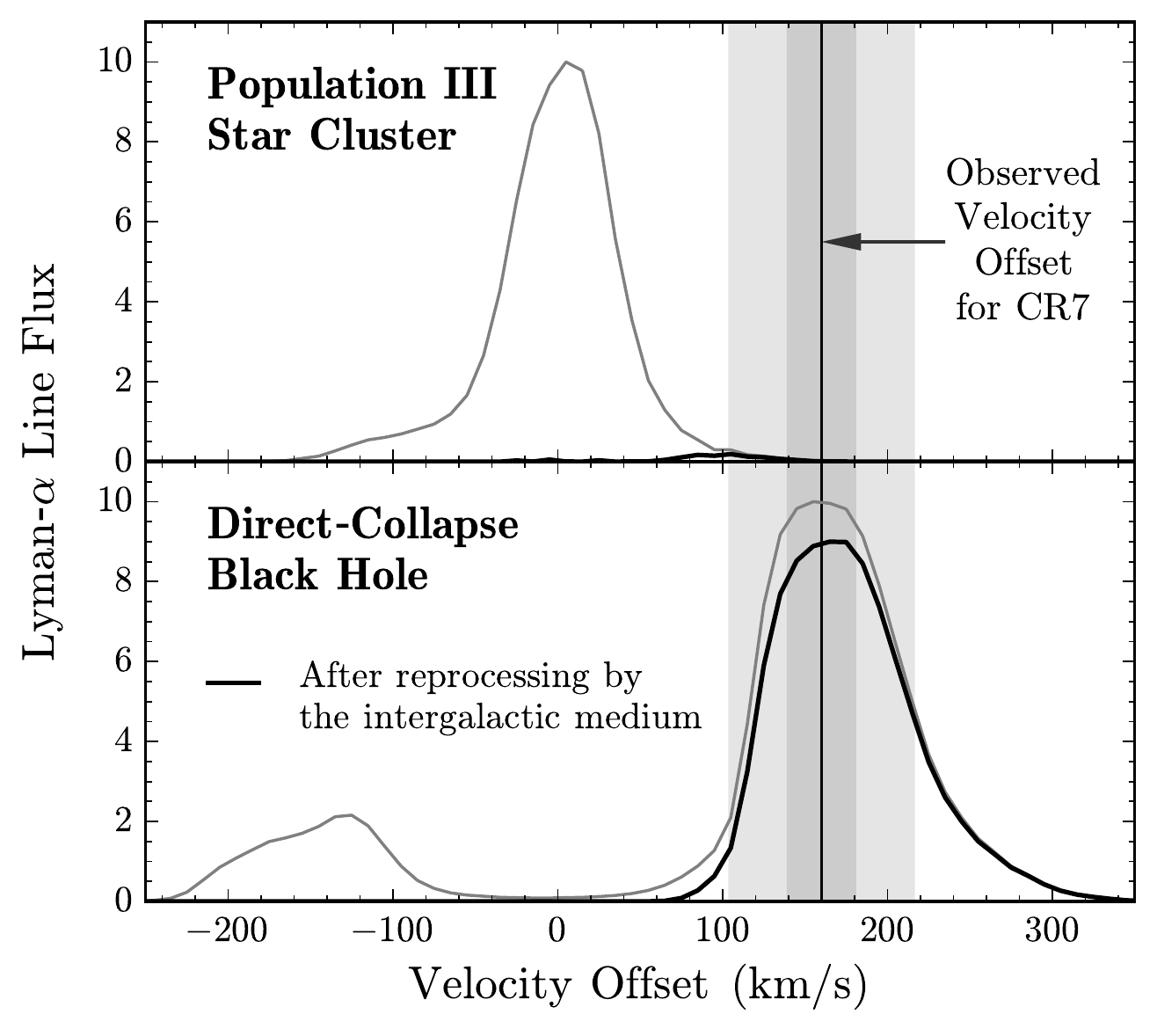}
    \caption{\protect\input{figures/CR7-flux/caption}}
    \label{fig:CR7-flux}
  \end{figure}

The DCBH scenario has received increased popularity as theoretical predictions agreed with multiple lines of observational evidence. Although individual cases currently remain tenuous, there is good reason to believe SMBH progenitor candidates may be observationally confirmed with the capabilities of next-generation observatories. Recently, the luminous COSMOS redshift~7~(CR7) Ly-$\alpha$ emitter at $z = 6.6$ was confirmed to have exceptionally strong Ly-$\alpha$ ($\gtrsim 8 \times 10^{43}\,\text{erg~s}^{-1}$) and possible \HeII\ 1640~\AA\ ($\sim 2 \times 10^{43}\,\text{erg~s}^{-1}$) emission with no detection of metal lines from the UV to the near-infrared within instrumental sensitivity \citep{Matthee_2015,Sobral_2015}. As a result, several groups have considered the CR7 source in the context of a young primordial starburst or DCBH \citep{Pallottini_2015,Agarwal_2016,Hartwig_2016,Visbal_CR7_2016,Dijkstra_DCBH_2016,Smidt_CR7_2016}. In \citet{Smith_CR7_2016}, we examined and reproduced several Ly-$\alpha$ signatures of the CR7 source under the Ly-$\alpha$ RHD framework discussed in the section above (see also figure~\ref{fig:CR7-cartoon}). As shown in figure~\ref{fig:CR7-flux}, the DCBH model reproduces the observed 160~km\,s$^{-1}$ velocity offset between the Ly-$\alpha$ and \HeII\ line peaks, whereas the stellar model fails. We also found that Ly-$\alpha$ radiation pressure turns out to be dynamically important in the case of CR7.

However, more recently, \citet{Bowler_CR7_2017} obtained deeper observations of CR7. 
The authors claim the new photometry cannot be reproduced by a DCBH spectral energy distribution (SED), suggesting instead that the broadband measurements may be contaminated by forbidden, doubly-ionized oxygen [\OIII] emission lines. They propose that CR7 can be classified as a more standard low-mass, narrow-line AGN or a low-metallicity starburst getting the hard SED from massive stellar binaries. In contrast, \citet{Pacucci_CR7_2017} argue that the new photometry is still consistent with the DCBH model. Either way, deep spectroscopy with future telescopes will be needed to discriminate convincingly between particular models \citep[see also][]{Agarwal_2017}. In the near future, other sources similar to CR7 may provide additional constraints on early galaxy and quasar formation. Indeed, \citet{Pacucci_DCBH_2016} identified two objects characterized by very red colours and robust X-ray detections in the CANDLES/GOODS-S survey with photometric redshift $z \gtrsim 6$ representing promising black hole seed candidates. We note that these objects were selected based on currently available \textit{Hubble Space Telescope} and \textit{Chandra Space Telescope} data.

Another independent argument for the existence of DCBHs is found in correlations between the cosmic infrared and X-ray backgrounds (CIB and CXB), which represent the cumulative light from faint, unresolved sources in the respective wavelength ranges \citep{Cappelluti_2013}. The specific signal is encoded within the source-subtracted CIB fluctuations after accounting for foreground stars and galaxies \citep[for additional details see][]{Kashlinsky_2005,Kashlinsky_2012}. Although other models may also explain the observations, DCBHs have been implicated as a natural way to produce enough IR and X-ray emission without over-ionizing the universe during the epoch of reionization \citep{Yue_2013,Helgason_2016}.

\subsection*{Outlook for the future}
Finally, we consider the prospects for unveiling the nature of supermassive black hole seeds with next-generation facilities. Continuing technological advances are going to allow us to probe the high-redshift universe in unprecedented detail. This includes the characterization of individual objects and integrated backgrounds based on observations in the radio, infrared, optical and X-ray wavelengths. Furthermore, space-based gravitational wave detectors will constrain SMBH merger models. Lastly, high-resolution simulations of black hole environments with multiscale physics will continue to refine our understanding of the formation and evolution of galactic black holes.

  \begin{figure*}
    \centering
    \includegraphics[width=\textwidth]{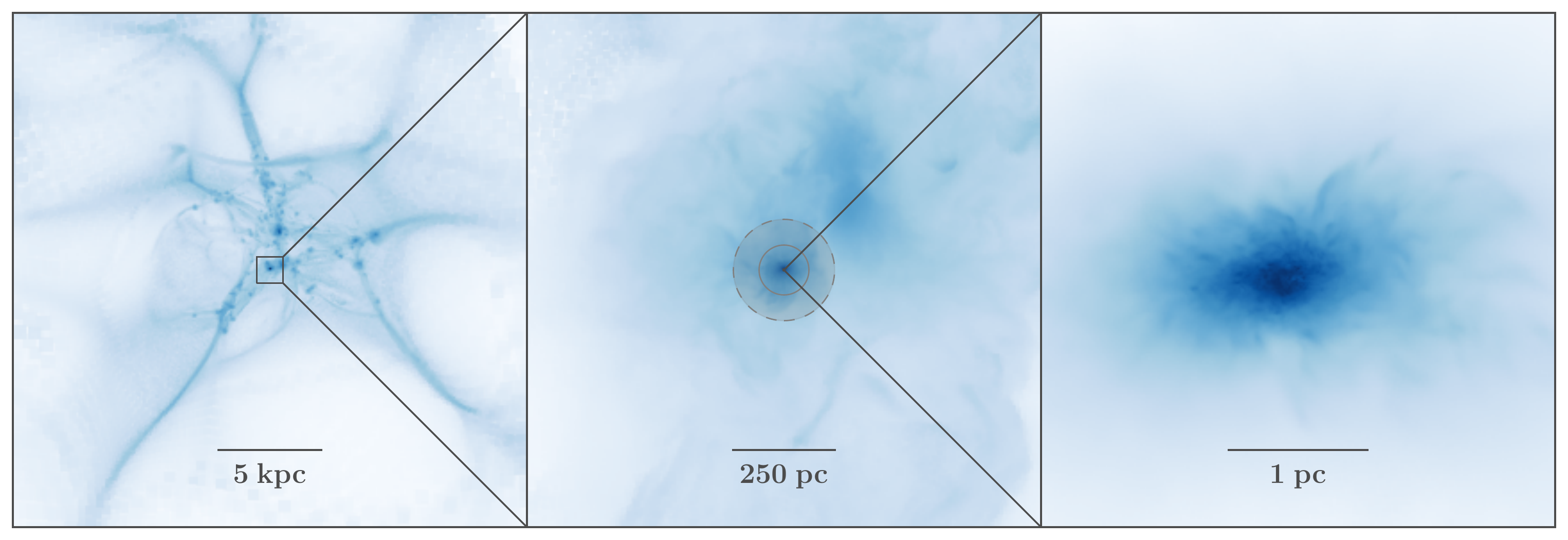}
    \caption{\protect\input{figures/density/caption}}
    \label{fig:density}
  \end{figure*}

The \textit{James Webb Space Telescope} (\textit{JWST}) and giant segmented mirror telescopes\ASfootnotemark[1]\footnotetext[1]{\;\,Telescopes with integral field spectrographs and adaptive optics imaging will include the Giant Magellan Telescope (GMT; \href{http://www.gmto.org/}{www.gmto.org}), Thirty Metre Telescope (TMT; \href{http://www.tmt.org/}{www.tmt.org}), and the European Extremely Large Telescope (E-ELT; \href{http://www.eso.org/sci/facilities/eelt/}{www.eso.org/sci/facilities/eelt}).} may allow direct observations of the first galaxies and quasars. In particular, the Near-Infrared Camera (NIRCam) and Near-Infrared Spectrometer (NIRSpec) instruments on board the \textit{JWST} will be capable of obtaining deep photometric and spectroscopic observations. The new data will yield a more complete census of galaxies from the first billion years of cosmic history, possibly including their redshifts, spectral energy distributions, star formation rates, metallicity, and emission line properties. As a reference, figure~\ref{fig:density} illustrates the gas density from an \textit{ab initio} cosmological simulation in which primordial gas undergoes the direct collapse to a black hole \citep{Becerra_2017}. We show the galaxy at different scales to highlight the filamentary large-scale structure, gas distribution and opaque cloud within a sub-parsec region.

Several other complementary observatories will contribute to the emerging picture as well. For example, black holes often produce jets with strong radio emission, which may be observed at higher redshifts with the Atacama Large Millimeter Array (ALMA\ASfootnotemark[2]). Upcoming 21-cm cosmology experiments, such as the Square Kilometre Array (SKA\ASfootnotemark[3]), will map the distribution of neutral hydrogen over the course of early cosmic history through reionization, providing a better understanding of the contribution of high-redshift quasars to this process. Future observations will also provide better measurements of the cosmic infrared and X-ray backgrounds which exhibit a correlation that might be explained by unresolved massive black holes in faint galaxies. Eventually, many of these sources will be resolved and characterized as the proposed Lynx and Athena X-ray telescopes detect high-energy emission from distant SMBHs. Deep surveys of nearby dwarf galaxies might also reveal traces of the black hole seeding mechanism due to their relative isolation after formation. Finally, the planned Laser Interferometer Space Antenna (eLISA\ASfootnotemark[4]) promises to directly detect these massive black holes via mergers, extending gravitational wave astronomy from the stellar-mass events recently detected with the Laser Interferometer Gravitational-Wave Observatory (LIGO\ASfootnotemark[5]) to the DCBH mass range.
\footnotetext[2]{\href{http://www.almaobservatory.org/}{www.almaobservatory.org}}
\footnotetext[3]{\href{http://skatelescope.org/}{skatelescope.org}}
\footnotetext[4]{\href{https://www.elisascience.org/}{www.elisascience.org}}
\footnotetext[5]{\href{http://www.ligo.org/}{www.ligo.org}}

\subsection*{Simulations}
The field of black hole research will benefit from the steady progress in computational algorithms and hardware, as high-resolution hydrodynamical simulations will provide additional insights into the process of SMBH formation. In particular, fully coupled radiation-hydrodynamics simulations will be crucial to elucidate the astrophysical phenomena responsible for the rapid growth of the first black holes. Pioneering simulations typically focus on specific aspects of larger questions in order to balance algorithmic complexity, resolution and computational feasibility. Eventually, it will be possible to apply more efficient and robust methods to problems with broader applicability. For example, incorporating effects based on accurate but traditionally expensive techniques, such as 3D Monte-Carlo radiative transfer, will be increasingly viable and worthwhile. The goal of these simulations is to connect what can be directly observed with what is ultimately powering these sources, however challenging this may be. A smoking-gun signature of an individual DCBH may be beyond the capabilities of next-generation telescopes, but the emergence of multiple independent lines of evidence might present a compelling picture in which massive black hole seeds bridge the gap in understanding the genesis of the first quasars.


\bibliographystyle{mnras}
\bibliography{biblio}

\label{lastpage}
\end{document}